\def\year{2021}\relax
\documentclass[letterpaper]{article} % DO NOT CHANGE THIS
\usepackage{aaai21}  % DO NOT CHANGE THIS
\usepackage{times}  % DO NOT CHANGE THIS
\usepackage{helvet} % DO NOT CHANGE THIS
\usepackage{courier}  % DO NOT CHANGE THIS
\usepackage[hyphens]{url}  % DO NOT CHANGE THIS
\usepackage{graphicx} % DO NOT CHANGE THIS
\usepackage{natbib}  % DO NOT CHANGE THIS AND DO NOT ADD ANY OPTIONS TO IT
\usepackage{caption} % DO NOT CHANGE THIS AND DO NOT ADD ANY OPTIONS TO IT
\usepackage{array}
\newcolumntype{L}[1]{>{\raggedright\let\newline\\\arraybackslash\hspace{0pt}}m{#1}}

\usepackage{booktabs}
\graphicspath{{./}{./figures/}}

\usepackage{lipsum}
\usepackage{todonotes}
\usepackage{xcolor}
\usepackage{amsmath}
\usepackage{amssymb}

\title{We Are in This Together: Quantifying Community\\ Subjective Wellbeing and Resilience }
\author{
    MeiXing Dong\textsuperscript{\rm 1},
    Ruixuan (Sophia) Sun*\textsuperscript{\rm 2},
    Laura Biester*\textsuperscript{\rm 1},
    Rada Mihalcea\textsuperscript{\rm 1} \\
}
\affiliations{
    \textsuperscript{\rm 1} Department of Computer Science, University of Michigan \\
    \textsuperscript{\rm 2} Department of Computer Science, University of Minnesota \\
    meixingd@umich.edu
}
\begin{document}

\maketitle

\begin{abstract}
The COVID-19 pandemic disrupted everyone's life across the world. 
In this work, we characterize the subjective wellbeing patterns of 112 cities across the United States during the pandemic prior to vaccine availability, as exhibited in subreddits corresponding to the cities. 
We quantify subjective wellbeing using positive and negative affect. We then measure the pandemic's impact by comparing a community's observed wellbeing with its expected wellbeing, as forecasted by time series models derived from prior to the pandemic.
We show that general community traits reflected in language can be predictive of community resilience.
We predict how the pandemic would impact the wellbeing of each community based on linguistic and interaction features from normal times \textit{before} the pandemic.
We find that communities with interaction characteristics corresponding to more closely connected users and higher engagement were less likely to be significantly
impacted. Notably, we find that communities that talked more about social ties normally experienced in-person, such as friends, family, and affiliations, were actually more likely to be impacted. Additionally, we use the same features to also predict how quickly each community would recover after the initial onset of the pandemic. We similarly find that communities that talked more about family, affiliations, and identifying as part of a group had a slower recovery.
\end{abstract}

\section{Introduction}
The COVID-19 pandemic has had widespread effects on people's subjective wellbeing. 
However, the local surrounding environment can greatly influence and be indicative of how people cope with an adverse event and shifting conditions. For instance, stronger social ties have been associated with higher wellbeing and community resilience\cite{kawachi2001social,aldrich2015social}. The aspects of life that a community gives attention to, such as leisure, family, and friends, can also be indicative of how that community may fare when impacted by a negative event.

Community resilience is a community's ability to adapt to, withstand, and recover quickly from changing conditions and disruptions \cite{berkes2013community,walsh2007traumatic}.
With the COVID-19 global pandemic,
boosting and maintaining mental wellbeing has become a prominent issue as everyone continues to grapple with the ongoing daily life restrictions and overall uncertainty of the situation. In this paper, we aim to understand the subjective affective wellbeing recovery patterns of communities in cities across the United States (US) and gain insight on relationships between cities' characteristics and the pandemic's effect on the cities' subjective wellbeing over time.

Many in-person interactions migrated online during the pandemic. Online forums such as Reddit offered a way for locals to stay connected and stay current with ongoing concerns such as whether certain restaurants were open or the status of vaccinations in their area. 
Reddit's city-focused subreddits, such as r/seattle and r/annarbor, correspond to cities across all states in the US. Discussions on these subreddits reflect people's concerns. 
While surveys could help measure wellbeing, they are often limited in scale. By looking at everyday conversational behavior in a community, we can get an aggregate picture of wellbeing. 

In our work, we characterize trends in how community subjective wellbeing shifted during the beginning of the COVID-19 pandemic (until the end of 2020) across 112 cities spread across the US. Using cities with similar wellbeing recovery patterns, we quantify what community characteristics correlate with lessened impact on wellbeing from the pandemic, as well as recovery speed given a negative impact. The features we examine are derived from online user interaction patterns and linguistic content.

We seek to answer the following research questions: 
\begin{enumerate}
    \item \textbf{How has the pandemic impacted the affective wellbeing of US cities?}
        
        We quantify wellbeing using positive and negative affect expressed in daily discussions on city-focused subreddits.
        We then measure the pandemic's impact on a city by comparing a city's observed wellbeing with its expected wellbeing, as forecasted by time series models derived from data prior to the pandemic. We define three patterns of wellbeing seen during 2020.
         
    \item \textbf{What distinguishes city communities that are more or less impacted by the pandemic?}
    
        We analyze a set of community traits derived from prior to the pandemic, encompassing linguistic features and user interaction patterns. We predict whether a city's wellbeing is heavily impacted by the onset of pandemic and analyze differences among cities that were more impacted and those that were not. 
        
    \item \textbf{How do impacted city communities differ in their speed of recovery?}
    
        Using the same community features, we predict whether an impacted city makes a recovery within the year, similarly analyzing distinguishing characteristics.
        
\end{enumerate}

\section{Related Work}
\textbf{Subjective Wellbeing.}
Research on subjective wellbeing (SWB) delves into how people feel and think about their lives \cite{Diener1999SubjectiveProgress}.
Subjective wellbeing is not a single concrete entity and studies look largely at two components, affective and cognitive wellbeing \cite{lucas1996discriminant}. Affective wellbeing is defined as the positive and negative emotions that people feel, such as happiness and anxiety. More positive affect and less negative affect indicates higher affective wellbeing.
On the other hand, cognitive wellbeing is defined as one's evaluation of one's life and resulting level of life satisfaction. This can refer to overall satisfaction, or satisfaction with respect to specific life domains such as jobs or relationships.   
These are two distinct constructs, and differ in many ways, such as their stability over time and how they are impacted by life events \cite{luhmann2012subjective}. Though affective wellbeing tends to return to baseline levels through hedonic adaptation \cite{frederick199916,lyubomirsky2011hedonic}, the adaptation rate can greatly differ for different people and circumstances \cite{Lucas2007AdaptationAT,luhmann2018hedonic}.

In our work, we present a longitudinal study of affective wellbeing and adaptation rates in different cities across the US in response to the COVID-19 pandemic, derived from large-scale naturally occurring social media data.

\textbf{Community Resilience and Social Capital.}
A community can be viewed as as a group of people who are bound by some common tie such as geographic location; they often have shared social norms \cite{Dong2019PerceptionsCultures}, values, and other characteristics. Constituent parts of a community can influence one another in complex ways. The study of community resilience investigates the qualities that allow a community to cope with and adapt to a collective disaster experience \cite{Ae2007CommunityReadiness}, in terms of physical and mental health outcomes. 

A core underlying concept is social capital \cite{Cutter2014,Sherrieb2010}. In the context of community resilience, social capital consists of the weak and strong social ties and networks in a community. This can come from social support, like family and friends, as well as social participation in the broader community. Improved connections can help provide support to members of the community.
    
In our work, we study the online counterparts of cities across the United States and how signals of social capital may influence their community wellbeing and resilience within the context of the COVID-19 pandemic.
        
\textbf{Studying Online Communities.}
Online communities have been the focus of research studying user behavior and community features. For instance, prior work has examined behavioral and linguistic factors that drive social support \cite{de2014mental,Ammari2019Self-declaredSupport,CunhaTheCommunity,andy-etal-2021-understanding} and signal community success \cite{Hamilton2017LoyaltyCommunities,Cunha2019AreCommunities}. A parallel line of work uses online communities as a lens to study the linguistic manifestation of mental health symptoms \cite{Fine2020,benton-etal-2017-multitask,coppersmith-etal-2014-quantifying}.
Work in this area \citep[e.g.,][]{kumar2015detecting,de2013predicting,pavalanathan2015identity,mitchell-etal-2015-quantifying} frequently makes use of Linguistic Inquiry and Word Count (LIWC) \citep{pennebaker2015development}, a tool that we use to measure city's affective wellbeing. With the onset of the COVID-19 pandemic, many have turned towards identifying the effects of the pandemic on mental health through the analysis of social media \cite{valdez2020social,biester2021covidmh}.

We adapt some of these characterizations of online communities and use them to distinguish between cities with different patterns of subjective wellbeing shifts during the COVID-19 pandemic. 
We measure community-level subjective wellbeing using social media, and draw insights from longer-term patterns of subjective wellbeing.

\section{Reddit City Communities}

\begin{table}[]
    \centering
    \begin{tabular}{llll}
        \toprule
        & Avg & Min & Max \\
        \midrule
        Authors & 21,717 & 1,799 & 116,338 \\
        Submissions & 17,104 & 1,506 & 80,231\\
        Comments & 349,756 & 9,021 & 2,490,517\\
        \bottomrule
    \end{tabular}
    \caption{Reddit city communities dataset statistics. Total values are computed per subreddit from all of 2017 - 2020, which are then aggregated over all subreddits. The corresponding cities cover 48 US states.}
    \label{tab:dataset_stats}
\end{table}

With the onset of the COVID-19 pandemic and corresponding stay-at-home measures, most of the socializing that normally happened in person shifted online. 
In our study, we focus on data collected from Reddit, a community-based online forum for information sharing and discussion. Sub-communities on Reddit, known as subreddits, can be centered on any interest. Some subbreddits have the intention of connecting people in a city and allowing them to discuss local news, events, attractions, and more. These allow us to study naturally occurring online counterparts to physical cities; from manual examination, we found that these communities appear to be largely composed of local residents who discuss city-related topics. 

During the COVID-19 pandemic, these subreddits became an important way for people to not only get local information, but also to express their worries with to who might best understand, and to find support.
Many posts are laden with emotion, such as fear or outrage towards policies being implemented, sadness when talking about a favorite business closing permanently, or loneliness when venting about social isolation and being unable to see friends and family.

Following prior work on subjective wellbeing \cite{Kramer2010AnHappiness,Mihalcea2006}, we examine the levels of positive and negative emotions expressed in language by members of the communities to gauge affective wellbeing in aggregate at the community level. We look at the posts and comments in these city subreddits to observe how affective wellbeing of communities in different US cities was affected over time by the COVID-19 pandemic.

We select the top five most populous cities from every US state, as well as Washington D.C., and manually find the corresponding subreddits. Not all cities have subreddits, yielding 233 cities.
For the cities with multiple subreddits, we choose the subreddit with the most members.
We also manually label each city with its corresponding county.

We collect all publicly available Reddit posts and comments from the 233 city subreddits starting from January 1st, 2017 to December 31st, 2020.\footnote{We used the Pushshift.io API.} We collect posts made prior to the pandemic as a way to compare with normal community behavior before the pandemic. Additionally, the period we study precedes widespread vaccine availability, allowing us to see how communities dealt with the pandemic when the main solutions involved limiting social interaction.

\begin{figure}
	\centering
	\includegraphics[width=\linewidth]{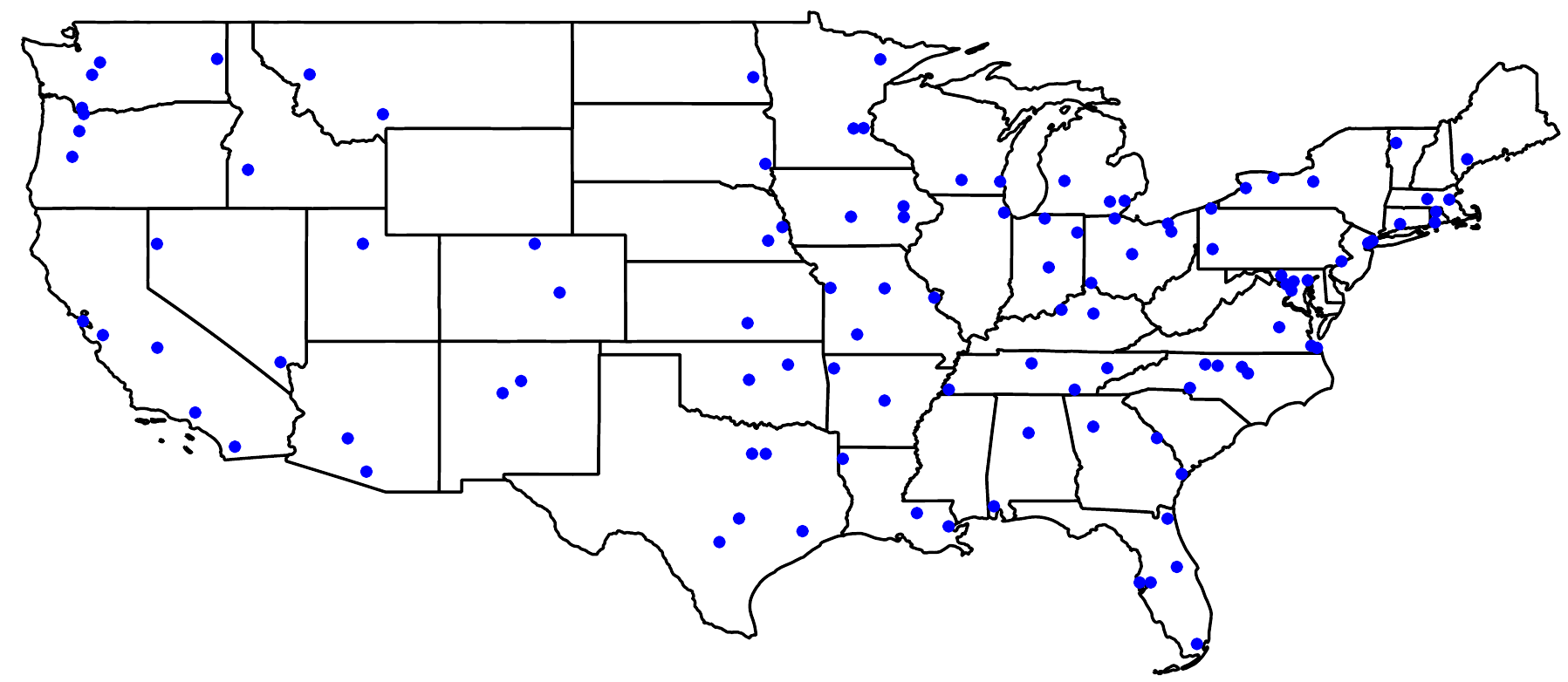}
	\caption{A map of the cities included in our city subreddit list, which are spread throughout the United States. Anchorage and Honolulu are also included in our city subreddits.}

	\label{fig:subreddit-map}

\end{figure}

To ensure that the cities have sufficiently active subreddits, we only use subreddits that have posts on at least 300 unique days in each of the years considered, resulting in a final list of 112 cities that cover 45 US states. 
We give statistics about the final dataset in Table \ref{tab:dataset_stats}, and a map of the cities is presented in Figure~\ref{fig:subreddit-map}.

\section{Quantifying Subjective Wellbeing and Resilience}
\label{sec:quantifying_wellbeing}

People can perceive the world in drastically different ways, even when they are experiencing similar events. How someone reacts to a traumatic or important situation can say a lot about how they are coping with the situation.
In particular, someone's emotional response is a crucial part of their reaction. We therefore focus on emotional expression as a way of quantifying mental wellbeing, which we define as a balance between the presence of positive emotion and a lack of negative emotion.

To compute wellbeing, we examine all posts of the members of a community and count the number of occurrences of words from the {\sc PosEmo} and {\sc NegEmo} categories of the Linguistic Inquiry and Word Count (LIWC2015) lexicon \cite{pennebaker2015development}. These correspond to positive emotion and negative emotion respectively. More formally, we define our metric for {\sc WellBeing} as:

$$\text{{\sc WellBeing}} = \text{{\sc PosEmo}}_{norm} - \text{{\sc NegEmo}}_{norm}$$

\noindent where $\text{{\sc PosEmo}}_{norm}$ and $\text{{\sc NegEmo}}_{norm}$ represent values after applying z-normalization to the average raw LIWC values for each day.

\subsection{Measuring Resilience} \label{sec:measuring_resilience}

Community resilience is a community's ability to adapt to, withstand, and recover quickly from changing conditions and disruptions.
From prior studies in psychology, we have seen that people tend to return to a baseline level of subjective wellbeing after life disruptions, even when the adverse situation persists \cite{luhmann2018hedonic}; 
if a community recovers and adapts more \textit{quickly}, then they are more \textit{resilient}. 

Based on this definition of resilience, we track the trend of {\sc WellBeing} scores during the pandemic. This trend can show how well the cities are coping with the pandemic, and if the cities recover (which we define as reaching expected {\sc WellBeing} scores predicted before the pandemic), it indicates strong resilience. On the other hand, if a city's {\sc WellBeing} decreases and does not recover for a long period, the city is likely having a harder time coping with the pandemic.

We begin by building a time series of {\sc WellBeing} values for each city subreddit. We first compute the average daily {\sc WellBeing} scores. Next, we fill in missing values using linear interpolation, and take a 7-day rolling mean for each day to smooth out weekly fluctuations and outliers.

We model the expected {\sc WellBeing} of a city's subreddit using the Prophet model \citep{taylor2018forecasting} trained on data prior to March 2020 (Jan. 1, 2017 - Feb. 29, 2020); this model has been used in prior work on the impacts of COVID-19 on social media forums \citep{biester2021covidmh,cao-etal-2021-quantifying}. The model fits the equation

\begin{equation}
    y(t) = g(t) + s(t) + \epsilon_t
\end{equation}

$g(t)$ represents a piecewise linear model that is used to represent the trend, while a Fourier series $s(t)$ is used to approximate thee yearly seasonality. The error is represented by the term $\epsilon_t$. We use the Prophet model trained on pre-COVID data to forecast post-COVID values through the end of 2020 (Mar. 1, 2020 - Dec. 31, 2020), along with the 95\% confidence interval.

We consider a city's {\sc WellBeing} to have significantly deviated below our expectations if the values for at least 25\% of the days fall below the 95\% confidence interval. We consider the early stage of the pandemic to be April 1 - June 30, 2020, and the middle stage of the pandemic to be July 31 - December 31, 2020. 
Based on these intervals, we define three classes to represent a city's resilience:
\begin{description}
    \item[Unaffected:] In the early stage of the pandemic, the city's {\sc WellBeing} does not significantly deviate below the expected values.
    \item[Recovered:] In the early stage of the pandemic, the city's {\sc WellBeing} significantly deviates below the expected values, but during at least one of the three-month periods in the middle stage, it no longer deviates.
    \item[Non-Recovered:] In the early stage of the pandemic, the city's {\sc WellBeing} significantly deviates below the expected values, and it continues to deviate during each of the three-month periods in the middle stage.
\end{description}

\begin{figure*}
	\centering
	\includegraphics[width=\linewidth, scale=0.8]{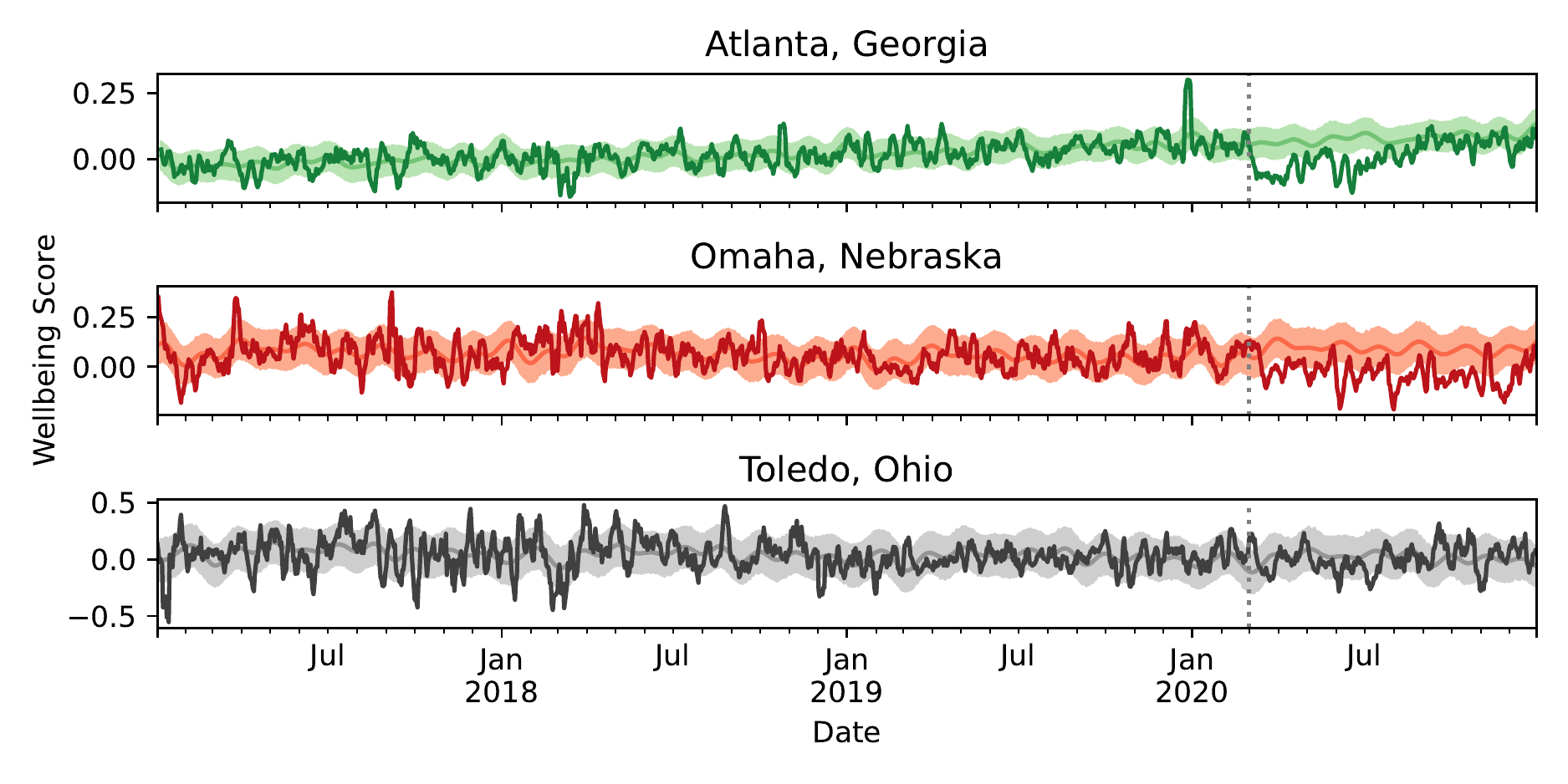}
	\caption{Average daily {\sc WellBeing} scores from a selection cities over time. Atlanta, Georgia \textit{recovered}, Omaha, Nebraska \textit{did not recover}, and Toledo, Ohio was \textit{unaffected}. The lighter line is the Prophet forecast, the shaded area is the 95\% prediction interval, and the darker line is the true value. The dotted line marks March 1st, 2020.}

	\label{fig:prophet-plot}

\end{figure*}

The number of cities matching each recovery pattern are shown in Table~\ref{tab:recovery_label_stats}. Figure~\ref{fig:prophet-plot} shows plots of the daily {\sc WellBeing} values for a selection of cities, which are used to determine their recovery patterns. In our experiments, we focus on two distinctions: (1) the distinction between cities that are affected by the pandemic (with respect to {\sc WellBeing} score), those that are not; and (2) the distinction between cities that recovered and those that did not.

\begin{table}[]
    \centering
    \begin{tabular}{ll}
        \toprule
        Recovery Pattern & Number of Cities\\
        \midrule
        Unaffected & 36 \\ % label 1
        Recovered & 32 \\ % labels 2: 12 and 3: 20
        Non-recovered & 44 \\ % label 4
        \textit{All} & \textit{112} \\
        \bottomrule
    \end{tabular}
    \caption{Number of cities that fall into each recovery pattern.}
    \label{tab:recovery_label_stats}
\end{table}

\subsection{Comparing with Traditional Resilience Metrics}
Community resilience, or the ability of a community to cope with a crisis, is a prominent topic in disaster and policy research \cite{Sherrieb2010}. Many have worked to define and quantify aspects of resilience, such as social capital and economic prosperity. Resilience metrics have been assigned to communities by analyzing empirical data about the community.  

One such well-known metric is Baseline Resilience Indicators for Communities (BRIC) \cite{Cutter2014}. BRIC measures the inherent resilience of counties in the US with respect to six different domains: social, economic, housing and infrastructure, institutional, community, and environmental.
The metric covers over 60 different variables, such as mental health support, educational attainment, and employment rate.\footnote{The data is from 2000 to 2013.} 
The raw data values are normalized using min-max scaling, and the orientation of values set so that larger values correspond theoretically to higher resilience, then combined to form  six different sub-indices corresponding to the different domains \cite{Cutter2014}. Resilience scores are assigned to each county in the US based on these measures.

We measure the Spearman correlation between our three classes of social media-derived resilience labels with the different components of the real-valued BRIC resilience scores. Each city is paired with its county's BRIC score; a county and score can be linked to multiple cities. The results are shown in Table \ref{tab:resilience_correlation}.

\begin{table}[]
    \centering
    \begin{tabular}{ll}
        \toprule
        BRIC domain & p-val \\
        \midrule
        Social & \textbf{0.004***}\\
        Economic & 0.680\\
        Housing and Infra & 0.554 \\
        Institutional & 0.142 \\
        Community & 0.664 \\
        Environmental & \textbf{0.018**} \\
        Aggregate & \textbf{0.070*} \\
        \bottomrule
    \end{tabular}
    \caption{Spearman correlation between the social media-derived resilience labels with the components of the BRIC resilience scores. Statistically significant values are bolded.\\ $***: p < 0.01$, $**: p < 0.05$, $*: p < 0.10$}
    \label{tab:resilience_correlation}
\end{table}

We see that there is correlation between our resilience labels and BRIC scores in a number of domains. Notably, we see high correlation with social resilience, as well as correlation with environmental and aggregate resilience. This demonstrates that our labels reflect existing measures of community resilience.

\section{City Community Features \label{sec:city_features}}
In order to build a model to predict whether a city's wellbeing will be affected and recover during COVID-19, we use a number of features, include demographic features (Section~\ref{sec:dem_features}), linguistic features (Section~\ref{sec:linguistic_features}), and interaction features (Section~\ref{sec:interaction_features}).

\subsection{Demographic Features}\label{sec:dem_features}
The demographic attributes of cities can be indicators of community resilience, as indicated by BRIC \cite{Cutter2014,walton2014csiro}. 
We examine a number of community demographic dimensions that may be related to city {\sc WellBeing} patterns during the pandemic.  

\textbf{Population Density.} Denser populations may provide more opportunities for people to aggregate and do things together, including unplanned interactions, therefore potentially facilitating the buildup of more social connections and social capital \cite{malecki2012regional}.

\textbf{Age.} Age is correlated with subjective wellbeing. For instance, in the US, subjective wellbeing tends to be lower during middle age and higher in younger and older adults \cite{Steptoe2015SubjectiveAgeing}. 

\textbf{Rent vs Own.} Areas where people own homes may be more invested in and connected to the surrounding community, resulting in stronger social bonds.

\textbf{Household Income.} Money can't buy happiness, but it can buy life necessities and provide stability. Higher income has been shown to be correlated with subjective wellbeing \cite{gardner2007money}. 

\textbf{Housing Cost.} Housing cost can be related to the overall quality of life in an area, although economic stress from housing cost could hamper subjective wellbeing. 

\textbf{Latitude.} Climate and weather can influence subjective wellbeing;
lower anxiety has been linked with higher temperatures and more hours of sunshine \cite{howarth1984multidimensional}. Higher-latitude regions, or areas further from the equator, tend to have colder temperatures and receive less light than lower-latitude areas.

Our demographic features are summarized in Table \ref{tab:city_dem_features}. We collect data from the U.S. Census Bureau\footnote{https://www.census.gov/data/developers/data-sets/acs-1year.html} and the National Weather Service\footnote{https://www.weather.gov/gis/Counties}, based on each city's corresponding Federal Information Processing Standard (FIPS) code.

\begin{table}[h!]
    \small
    \centering
    \begin{tabular}{ >{\raggedright}p{2cm}p{5cm}  }
     \toprule
     Feature Name & Description\\
     \midrule
     {\sc Population Density} & Number of people per square mile of land area \\
     \midrule
     {\sc Median Age} & Median age of residents of the city\\
     \midrule
     {\sc Rent vs. Own} & The ratio of number of people who rent a property to those who own a property \\
     \midrule
     {\sc Median Household Income} & Median household income in the past 12 months \\
     \midrule
     {\sc Median Housing Cost} & Median monthly housing cost \\
     \midrule
     {\sc Latitude} & The latitude of the county where a city is located \\
     \bottomrule
     
    \end{tabular}
    \caption{Summary of city demographic features. All values are derived from 2019, prior to the pandemic.}
    \label{tab:city_dem_features}
\end{table}

\subsection{Covid Related Features}

\textbf{Cases and Deaths.}
Local COVID-19 case counts and deaths measure the direct impact of the disease itself on members of the community. As a result, cities with more cases and deaths early on in the pandemic may be more heavily impacted. We collect county-level case and death counts from The New York Times, based on reports from state and local health agencies \cite{nytcoviddata}. We consider average daily {\sc Cases} and {\sc Deaths} per person in the early, middle, and late periods defined in Section~\ref{sec:measuring_resilience}.

\textbf{Mask Mandates.} Mask usage was a significant social behavior change during the pandemic. We gather county-level daily mask mandate data\footnote{https://data.cdc.gov/Policy-Surveillance/U-S-State-and-Territorial-Public-Mask-Mandates-Fro/62d6-pm5i/data} from April 10, 2020 through August 15, 2021 based on county FIPS codes. We consider the ratio of days that a mask mandate was in place within each time period ({\sc Mask Mandate Percentage}).

\subsection{Linguistic Content Features} \label{sec:linguistic_features}

\textbf{Personal Concerns.}
%- FAMILY, FRIEND, WORK, LEISURE, HOME, HEALTH \\
The sharing of personal concerns and details about personal life can garner social support online \cite{Choudhury2017TheRisk}. People may discuss a wide range of personal attributes, such as their family, career, and hobbies.
To quantify people's discussion of personal concerns, we use the LIWC lexicon \citep{pennebaker2015development} and consider the following categories: \textit{family}, \textit{friend}, \textit{work}, \textit{leisure}, \textit{home}, and \textit{health}. 

\textbf{Group Belonging.} 
When people talk frequently about their affiliations, such as work associates, romantic partners, or neighbors, it can indicate a sense of community and presence of social ties. Social ties have been shown to be beneficial for wellbeing \cite{kawachi2001social}.
Similarly, people can also indicate a sense of group membership through the use of pronouns like ``we'' or ``our.'' We therefore consider the LIWC categories \textit{affiliation} and \textit{we}.

\textbf{Time Orientation.}
%- FOCUSFUTURE, FOCUSPAST, FOCUSPRESENT \\
One's time orientation is the emphases that one places on each of three relative time periods: the past, present, and future \cite{wallace1960temporal}. This can be indicative of one's subjective well-being and mental health. For instance, people who focus on the past in a positive way can find joy in their memories \cite{zimbardo2015putting}. 
We consider three measures of time perspective and orientation: \textit{past focus}, \textit{future focus}, and \textit{present focus}. All three are computed using the corresponding categories in LIWC: {\sc Focus Past}, {\sc Focus Future}, and {\sc Focus Present}. The {\sc Focus Past} category primarily consists of past-tense verbs, which also reveal how much people share about their activities.

In our experiments, we derive a set of values for each city corresponding to how prevalent each of these categories were in their subreddits during normal times prior to the pandemic. For each of the lexicon features, we compute the percentage of the category present in each post and average over all posts in 2019. 

\subsection{Pragmatic Features.} We consider two pragmatic features that build on the lexical features we extract using LIWC.

\textbf{Empathy.} The first is empathy, which Empathy is a positive attribute of online conversations, and influences people's willingness to open up about personal experiences \cite{zhao2013trust}. We measure this using a BERT-based model \cite{zhou-jurgens-2020-condolence}. 

\textbf{Toxicity.} Anti-social behavior can hurt the sense of community that people feel online. We measure toxicity using the Perspective API\footnote{\url{https://www.perspectiveapi.com/}}
\cite{delise2018toxicity,vidgen-etal-2020-recalibrating}. Due to the rate limitation for the Perspective API, we sample 18,250 posts (equivalent to 50 posts per day) from 2019 per subreddit, or all posts if the subreddit has fewer posts. We take the mean toxicity score across these posts.

\subsection{Interaction Features}\label{sec:interaction_features}
We hypothesize that the way in which people interact with others on their subreddit may be indicative of the community strength and resilience in the face of a disaster event such as the COVID pandemic. In order to measure interactions, we compute \textit{user interaction features},  which represent how individual users interact with one-another; and \textit{post interaction features}, which represent the interactions with user's \textit{posts}. The metrics expand on those used in \citet{biester2021covidmh} which were drawn from \citet{tang2010graph, wilson2009user, williams2015network, kang2017semantic}. 

\textbf{User Interaction Features.} \label{sec:graph-features}
The user interaction features are computed based on a graph representing daily interactions between users, where an interaction occurs when user $B$ comments on user $A$'s post or comment. Doing so introduces an edge from $B$ to $A$ and vice-versa, as we use an undirected graph. For each day, we consider the posts and comments that occurred on that day and create edges to their parents (regardless of when they occurred). Next, we compute the graph metrics that are shown in Table~\ref{tab:graph-metrics} for each graph, using the NetworkX package \cite{SciPyProceedings_11}. When computing our user interaction features for classification, we compute the average over all of the days where there is activity in the subreddit for each metric.

The metrics represent many facets of the community structure; node count and edge count reflect the daily activity on the subreddit. Density represents how connected each member of the community is to all other members; mean eccentricity (the maximum distance of each vertex from any other other vertex) represents a similar phenomena. Connected component count represents the number of distinct sub-groups in the community who do not interact on a given day, while mean connected component counts represent the size of those subgroups. Mean shortest path (across all components) and diameter represent the distance between nodes that are directly or indirectly connected.

\textbf{Post Interaction Features} \label{sec:tree-features}
To measure how users interact with posts in each subreddit, we first form a tree for each post that represents the chain of comments created in response to it. In each tree, the post is the root node and each comment is a child of the post or comment it replied to. For each post, we compute five measures shown in Table~\ref{tab:post-interaction-measures}, some of which are attributes of its corresponding tree and others which the timing of its comments. The measures are inspired by prior work on social media interaction \cite{wei-etal-2016-post,de2016discovering}. The measures are averaged across all posts in the subreddit made during 2019 to compute our final features. We began with a longer list of features and removed many features that were highly correlated with our final set. 

The first two metrics, {\sc Tree Size} and {\sc Direct Reply Count} are representative of the number of comments a post receives. {\sc Leaf Node Count} represents the number of comments that are left without a reply, while {\sc Max Level Width} represents the number of comments at the largest level of the tree. The final metric represents how long on average it takes for each post to get a comment. When a metric would otherwise be undefined, we only consider posts that have comments.

\begin{table}[h!]
    \small
{\begin{tabular}{ >{\raggedright}p{2.5cm}p{5cm} }
    \toprule
   Feature Name  &  Description \\ \midrule
    {\sc Node Count $|N|$} & Number of unique users who posted or commented  \\ \midrule
    {\sc Edge Count $|E|$} & Number of unique users who interacted through a reply to a post or comment \\ \midrule% \hline
    {\sc Mean Degree} & Mean number of edges per node \\ \midrule %\hline
    {\sc Density $\frac{2|E|}{|N|(|N| - 1)}$} & Number of edges in graph over number of possible edges \\ \midrule %\hline
    {\sc CC Count} & Number of subgraphs in which all pairs of nodes are connected by an edge \\  \midrule%\hline
    {\sc Mean Eccentricity} & Mean of eccentricity across nodes; eccentricity is the maximum distance between node $n$ and any other node\\ \midrule %\hline
    {\sc Mean CC Size} & Mean number of nodes in a connected component \\ \midrule %\hline
    {\sc Mean Shortest Path } & Mean distance between each pair of vertices that are connected by a path\\ \midrule %\hline
    {\sc Diameter} & Maximum distance between any pair of nodes within a connected component \\
    \bottomrule %\hline
\end{tabular}}
\caption{Summary of user interaction features. All values are derived from 2019, prior to the pandemic.}
\label{tab:graph-metrics}
\end{table}

\begin{table}
\small
{\begin{tabular}{ >{\raggedright}p{2.5cm}p{5cm}  }
    \toprule
  Feature Name  &  Description \\
    \midrule
    {\sc Tree Size} & Number of nodes in tree \\
    \midrule
    {\sc Direct Reply Count} & Number of children of the head node\\
    \midrule
    {\sc Leaf Node Count} & Number of leaf nodes in the tree \\
    \midrule
    {\sc Max Level Width} & Number of nodes in the largest level of the tree \\
    \midrule
    {\sc Min Response Time} & Time between creation of original post and the first comment it received \\
    \bottomrule
\end{tabular}}
\caption{Summary of post interaction features. All values are derived from 2019, prior to the pandemic.}
\label{tab:post-interaction-measures}
\end{table}

\section{Predicting Cities' Pandemic Impact} 
A community's normal behavior is predictive of how they may cope with an adverse event. For instance, communities with stronger social ties may be more supportive of community members during a disaster, leading to higher subjective wellbeing. Similarly, the general disposition of community members, such as being future oriented or valuing leisure activities, may also indicate coping patterns.

We examine community interaction and linguistic characteristics of city subreddits during normal times prior to the pandemic, in 2019, as previously described in Section \ref{sec:city_features}. For the COVID-related features, only the early time period is used here, as there can be no causal relationship between the middle and late periods and wellbeing in the early months of the pandemic. Using these features, we build models to predict whether a city's subjective wellbeing will be significantly negatively impacted by the onset of the COVID-19 pandemic. We distinguish between cities that are unaffected and all those that are affected, as exhibited in the first few months following the start of the pandemic (Section  \ref{sec:quantifying_wellbeing}).

We train a logistic regression classifier with leave-one-out cross validation (LOO-CV) across all the cities. The features are scaled such that they all take on values from 0 to 1. 
Given the greater prevalence of affected cities and limited number of city data samples, we balance our training data by oversampling the minority class of unaffected cities using SMOTE \cite{chawla2002smote}.

\subsection{Results and Discussion}

We show our results for distinguishing affected versus unaffected cities in Table \ref{tab:binary_no_change_v_all_results}. All metrics other than accuracy are computed using macro averaging.

We see that all of our COVID-related, community interaction and linguistic feature sets exceed the random baseline of 50\%. Further, user interaction features perform the best.

Additionally, we examine how different features correlate with community resilience by analyzing the coefficients of our regression model that is trained using all of the features together. We use the mean coefficient values across all folds in the LOO-CV. The results are listed in Table~\ref{tab:binary_no_change_v_all_coef}.

\begin{table}[h!]
\centering
\begin{tabular}{l|lllll}
\toprule
     Feature Set &            Acc &              P &              R &             F1 &            AUC \\
\midrule
          Random &          0.500 &            --- &            --- &            --- &            --- \\
    Demographics &          0.607 &          0.601 &          0.615 &          0.591 &          0.693 \\
   COVID-Related &          0.696 &          0.645 &          0.630 &          0.635 &          0.603 \\
User Interaction &          0.741* &          0.718 &          0.743 &          0.722 &          0.801 \\
Post Interaction &          0.661 &          0.638 &          0.655 &          0.638 &          0.698 \\
       Pragmatic &          0.688 &          0.668 &          0.689 &          0.668 &          0.696 \\
            LIWC &          0.688 &          0.663 &          0.682 &          0.665 &          0.716 \\
    \textit{All} & \textbf{0.768*} & \textbf{0.744} & \textbf{0.770} & \textbf{0.750} & \textbf{0.823} \\
\bottomrule
\end{tabular}
\caption{Unaffected vs. Affected classification results. $* : p < 0.05$, when compared with Demographics features, using a t-test; significance tests performed only for accuracy (Acc). }
\label{tab:binary_no_change_v_all_results}
\end{table}

\textbf{Demographic Features.}
Unaffected cities are associated with higher {\sc Latitude}. Though higher latitudes have colder weather which has previously been correlated with lower wellbeing in general, people living there may also be more resilient since they are used to dealing with some amount of discomfort throughout much of the year. These cities also are associated with higher {\sc Population Density}, and therefore more people and social ties. They also are likely to have a higher {\sc Rent vs Own} ratio, and higher {\sc Median Housing Cost}, which likely is tied to the higher population density, as most people in higher density cities rent and also have a higher cost of living. Finally, these cities are associated with a higher {\sc Median Age}, which may indicate more stable and invested long-term occupants.

Affected cities are associated with higher household income, which is counterintuitive as higher income often correlates with higher subjective wellbeing. However, since we are looking at resilience to wellbeing impact and not absolute wellbeing, it may be that areas with higher income are less accustomed to large negative events and their affective wellbeing was more impacted.

\begin{table}[h!]
\centering
\begin{tabular}{l|p{4cm}r}
\toprule
     Feature Set &                       Feature Name &   Coef \\
\midrule
    Demographics &                     \sc{ Latitude} &  1.187 \\
   COVID-Related &                    \sc{Early Mask} &  1.095 \\
   COVID-Related &                   \sc{Early Cases} &  0.810 \\
   COVID-Related &                  \sc{Early Deaths} &  0.741 \\
User Interaction &                       \sc{Density} &  0.725 \\
       Pragmatic &                      \sc{Toxicity} &  0.620 \\
    Demographics &           \sc{ Population Density} &  0.616 \\
    Demographics &   \sc{Median Housing Cost} &  0.410 \\
            LIWC &                 \sc{Focus Present} &  0.352 \\
            LIWC &                            \sc{We} &  0.312 \\
    Demographics &              \sc{Rent vs Own Rate} &  0.267 \\
Post Interaction &                     \sc{Tree Size} &  0.175 \\
Post Interaction &            \sc{Direct Reply Count} &  0.172 \\
    Demographics &                    \sc{Median Age} &  0.149 \\
            LIWC &                    \sc{Focus Past} &  0.136 \\
Post Interaction &               \sc{Max Level Width} &  0.130 \\
Post Interaction &               \sc{Leaf Node Count} &  0.093 \\
            LIWC &                          \sc{Home} &  0.074 \\
User Interaction &                    \sc{Edge Count} & -0.010 \\
User Interaction &                   \sc{Mean Degree} & -0.073 \\
            LIWC &                   \sc{Affiliation} & -0.089 \\
User Interaction & \sc{Mean CC Size} & -0.117 \\
User Interaction &                    \sc{Node Count} & -0.210 \\
       Pragmatic &                       \sc{Empathy} & -0.213 \\
            LIWC &                  \sc{Focus Future} & -0.241 \\
            LIWC &                        \sc{Friend} & -0.371 \\
            LIWC &                          \sc{Work} & -0.432 \\
Post Interaction &             \sc{Min Response Time} & -0.462 \\
            LIWC &                       \sc{Leisure} & -0.493 \\
            LIWC &                        \sc{Family} & -0.526 \\
    Demographics &              \sc{Household Income} & -0.543 \\
User Interaction &     \sc{CC Count} & -0.606 \\
            LIWC &                        \sc{Health} & -1.168 \\
User Interaction &             \sc{Mean Eccentricity} & -1.250 \\
User Interaction &            \sc{Mean Shortest Path} & -1.504 \\
User Interaction &                      \sc{Diameter} & -1.519 \\
\bottomrule
\end{tabular}
\caption{Unaffected vs. Affected coefficients. Positive coefficients indicate that the feature is more associated with the subreddits of cities that are unaffected; negative coefficients indicate association with the affected subreddits of cities.}
\label{tab:binary_no_change_v_all_coef}
\end{table}

\textbf{Covid Related Features.} Unaffected city subreddits are more associated with all three COVID-related features, including mask mandates, cases, and deaths. This somewhat surprising result indicates that low rates of COVID still led to a dip in wellbeing as people navigated new routines. Legislative decisions with major impacts such as school closures may have led to additional negative emotion in areas where case counts and deaths were low.

\textbf{User Interaction Features.} Subreddits where {\sc WellBeing} was not highly impacted by the pandemic are associated with higher density, meaning that their members tend to interact more with others.

Affected subreddits are associated with more separate groups of connected users ({\sc CC Count}), meaning that individuals are not interacting as much with the larger community, and that interactions are more localized. These subreddits are also associated with being less connected, with higher {\sc Eccentricity}, longer {\sc Mean Shortest Path}, and larger {\sc Diameter}.

\textbf{Post Interaction Features.}
Cities whose {\sc WellBeing} is not impacted by COVID-19 tend to have posts with more interaction, as shown by correlation with greater {\sc Tree Size}, {\sc Direct Reply Count}, {\sc Max Level Width}, and {\sc Leaf Node Count}.
%- tree size, direct reply count, max level width, leaf node count
On the other hand, affected cities are correlated with higher {\sc Min Response Time}.

This indicates that subreddits with \textit{slower} response times and \textit{less} interaction end up with greater {\sc WellBeing} impact during the first three months of the COVID pandemic.

\textbf{Linguistic Features.}
Unaffected cities are correlated with the {\sc Focus Past}, {\sc Focus Present}, and {\sc Home} LIWC categories. The {\sc Focus Past} and {\sc Focus Present} categories are largely verbs in the present and past tense, and therefore this may indicate that people are more willing to share about the activities they do, and details about their lives overall. People talking about their home more, prior to the pandemic, may indicate that their home environment is an important part of their normal life. Therefore, when the pandemic hit and people were largely confined to their homes, this may have not been as significant of a negative shock. 

We see that affected cities are more likely to talk about their friends, family, and other affiliations. Though many of these would generally be positive social factors in a disaster, the nature of the social policies put in place may have turned these into negative factors, since social interaction was specifically restricted. 

We also see that the affected cities are more likely to focus on the future. Similarly, while this might normally mean hopefulness towards the future, it also may mean result in greater wellbeing impact from the pandemic, since the state of the world was under such great uncertainty.

People generally talking about their health more may mean they had more health concerns, which could have been exacerbated by avoiding the hospital and also being more concerned in general about health issues.

\textbf{Pragmatic Features.}
Finally, toxicity is associated with unaffected cities while empathy is associated with affected cities. This may indicate that people opened up about the struggles of the pandemic when expecting empathetic rather than toxic responses from the online community.

\section{Predicting Cities' Ability to Recover}
% Non-Recovered vs. Recovered
We now look at distinguishing between those cities that recovered by the end of 2020 versus those that had not, among those that were affected. We maintain the same experimental setup as in the previous section.

\subsection{Results and Discussion}

\begin{table}
\centering
\begin{tabular}{l|lllll}
\toprule
     Feature Set &            Acc &              P &              R &             F1 &            AUC \\
\midrule
          Random &          0.500 &            --- &            --- &            --- &            --- \\
    Demographics &          0.382 &          0.392 &          0.393 &          0.381 &          0.392 \\
   COVID-Related &          0.526 &          0.557 &          0.553 &          0.523 &          0.395 \\
User Interaction &          0.461 &          0.482 &          0.483 &          0.458 &          0.479 \\
Post Interaction &          0.447 &          0.433 &          0.433 &          0.433 &          0.317 \\
       Pragmatic &          0.553* &          0.549 &          0.550 &          0.548 &          0.488 \\
            LIWC &          0.539* &          0.542 &          0.543 &          0.537 &          0.562 \\
    \textit{All} & 0.566* & 0.572 & 0.574 & 0.565 &          0.546 \\
    
    \textit{All (selected)} & \textbf{0.605*} & \textbf{0.601} & \textbf{0.604} & \textbf{0.601} & \textbf{0.565} \\
\bottomrule
\end{tabular}
\caption{Non-Recovered vs. Recovered classification results. All (selected) only includes feature sets that were better than random: COVID-Related, Pragmatic, and LIWC. $*: p < 0.05$, when compared with Demographics features, using a t-test; significance tests performed only for accuracy (Acc). }
\label{tab:binary_exclude_no_change_results}
\end{table}

We show the classification results in Table \ref{tab:binary_exclude_no_change_results}. We see that the task of distinguishing between recovered and non-recovered cities is a more difficult task; only the pragmatic, LIWC, and COVID-related features are able to predict recovery better than random chance. Therefore, our analysis here focuses only on these features. Logistic regression coefficients from a model using only these feature sets are shown in Table \ref{tab:binary_exclude_no_change_coef}.

\textbf{COVID Related Features.} The COVID-related features indicate that cities with mask mandates, even if they still have a large number of cases, were likely to have recovered by late 2020, indicating an ability to cope with the pandemic prior to vaccines. Larger death counts correspond with non-recovered cities, although with very small coefficients.

\textbf{Linguistic Features.}
We see that recovered cities are more associated with past-tense language. As noted before, this likely indicates people sharing more about what they have done, as the {\sc Focus Past} LIWC category consists primarily of past-tense verbs. They also talk more about their friends and homes.

On the other hand, non-recovered cities talk more about family, affiliations, and work. These may have been large aspects of life for people in these cities, which were then greatly impacted by the pandemic due to social distancing policies keeping people from seeing their extended families and co-workers. They also refer to themselves with {\sc We} words more, indicating that they considered themselves affiliated with groups. This feeling of connection was likely impacted by the isolation during the pandemic.

\textbf{Pragmatic Features.}
These reveal that more toxic communities are less likely to have recovered; this may be due to overall negative sentiment and disagreement in the communities about COVID-related measures. Interestingly, empathetic communities are also less likely to be recovered, but perhaps due to people being more open about their troubles.

\begin{table}[h!]
\centering
\begin{tabular}{l|lr}
\toprule
  Feature Set &       Feature Name &   Coef \\
\midrule
         LIWC &    \sc{Focus Past} &  0.981 \\
COVID-Related &     \sc{Late Mask} &  0.630 \\
         LIWC &        \sc{Friend} &  0.489 \\
COVID-Related &    \sc{Late Cases} &  0.332 \\
COVID-Related &  \sc{Middle Cases} &  0.265 \\
         LIWC &          \sc{Home} &  0.251 \\
         LIWC & \sc{Focus Present} &  0.158 \\
         LIWC &  \sc{Focus Future} &  0.145 \\
COVID-Related &   \sc{Middle Mask} &  0.099 \\
COVID-Related &    \sc{Early Mask} & -0.005 \\
COVID-Related &   \sc{Early Cases} & -0.049 \\
         LIWC &       \sc{Leisure} & -0.052 \\
COVID-Related &  \sc{Early Deaths} & -0.060 \\
COVID-Related & \sc{Middle Deaths} & -0.079 \\
COVID-Related &   \sc{Late Deaths} & -0.090 \\
    Pragmatic &       \sc{Empathy} & -0.300 \\
         LIWC &        \sc{Family} & -0.387 \\
         LIWC &   \sc{Affiliation} & -0.409 \\
    Pragmatic &      \sc{Toxicity} & -0.455 \\
         LIWC &        \sc{Health} & -0.476 \\
         LIWC &            \sc{We} & -0.574 \\
         LIWC &          \sc{Work} & -0.683 \\
\bottomrule
\end{tabular}
\caption{Non-Recovered vs. Recovered coefficients. Positive coefficients indicate that the feature is more associated with the subreddits of cities that recovered; negative coefficients indicate association with the subreddits of cities that did not recover. Only COVID-related, pragmatic, and LIWC features are included, as other features did not yield performance exceeding random chance.}
\label{tab:binary_exclude_no_change_coef}
\end{table}

\textbf{Demographic and Interaction Features.}
We find that demographic features are the least predictive, in line with our results from the first task, showing that the behavior of the city is not easily explained by city characteristics such as size or income. We also find that the interaction features are less predictive while the LIWC and pragmatic features do better, implying that \textit{what} people discuss and \textit{how} people communicate are more indicative of longer-term recovery patterns as compared to patterns of social connections.

\section{Broader Implications and Ethical Considerations}
Our work, in conjunction with existing work \cite{ashokkumar2021covidcities,biester2021covidmh}, shows that the pandemic had different effects on different communities. Further, we show that signals from social media can be predictive of how a community copes with adversity.
We found that cities more affected by the pandemic tended to have less connected members and had previously placed more importance on life aspects that were most impacted by social distancing during the pandemic, such as seeing friends and participating in group activities. Our features were predictive of whether a city's wellbeing was affected by the pandemic. However, predicting the subsequent recovery trajectory of affected cities proved to be more difficult, implying that there are other factors involved and further work is needed to understand community resilience over time.

Our findings indicate that differential policies should be put in place for communities, based on the pandemic's local impact.
Cities more impacted by social distancing measures may need to place higher priority on re-establishing social activities, such as local events and cultural festivals. 
Furthermore, policymakers could use automatically-derived signals from social media as a real-time source of feedback for their policies, especially during times that require quick decision-making like during the pandemic. However, it's important that such factors are considered holistically.

A limitation of our findings is that they do not necessarily reflect the general public. Our work focuses on a single social media site (Reddit) where the users tend to be young\footnote{https://www.statista.com/statistics/261766/share-of-us-internet-users-who-use-reddit-by-age-group/} and male.\footnote{https://www.statista.com/statistics/1255182/distribution-of-users-on-reddit-worldwide-gender/} Further, Reddit activity may potentially overrepresent the number of people affected by the pandemic (e.g., more people joining and posting due to physical lockdowns) or exclude those who largely ignored the pandemic. Our analyses look at \textit{all} posts and do not focus only on posts that relate to COVID-19, which may lessen this bias. Additionally, there are many other facets of how communities acted during COVID, such as their general compliance with COVID-19 prevention measures or vaccination rates. However, further studies using other social media, surveys, and more facets of how communities acted during COVID (e.g., mask adherence, general compliance with COVID-19 prevention measures) can help support and extend our insights.

Because our study is based solely on observational data, we cannot establish causal links between the community characteristics we have identified and the wellbeing recovery outcomes. To address this, future work could involve collecting ground truth data about city recovery and resilience, such as through large surveys of individuals in each city regarding their wellbeing during the pandemic.

Finally, our study should not be construed to be a comprehensive study of wellbeing.
Subjective wellbeing does not consist solely of the presence of positive affect. It is more complex and multi-faceted, involving other aspects such as life satisfaction which are impacted in different ways \cite{kettlewell2020differential}. Future work could study how community factors relate to these additional aspects of wellbeing. Furthermore, the relation of our wellbeing metric to metrics such as self-reported life satisfaction has not been studied; a misalignment between stance expressed in social media and public opinion surveys has been noted in prior work \cite{joseph-etal-2021-mis}, and we leave it to future work to study how wellbeing as expressed in social media posts relates to self-reported wellbeing.

\section{Conclusion}
We characterized the affective wellbeing patterns of cities across the US during the COVID-19 pandemic prior to vaccine availability, as exhibited in subreddits corresponding to the cities. We then derived linguistic and interaction features from the subreddit communities based on data prior to the pandemic and used them to predict how the affective wellbeing of each community would be impacted by the pandemic. We showed that communities with interaction characteristics corresponding to more closely connected users and higher engagement were less likely to be significantly impacted. Notably, we found that communities that talked more about social ties, such as friends, family, and affiliations, were actually more likely to be impacted. This may result from the social isolation policies affecting precisely these social ties.
Additionally, we used the same features to predict how quickly each community would recover after the initial onset of the pandemic. We similarly found that communities that talked more about family, affiliations, and identifying as part of a group were more likely to recover more slowly. We showed that general community traits can be predictive of community resilience.
%\begin{acks}
%To Robert, for the bagels and explaining CMYK and color spaces.
%\end{acks}

\bibliographystyle{aaai21}
\bibliography{references/references, references/references-sophia, references/references-laura-covid-paper, references/references-meixing}

\end{document}